\documentclass{appolb}
\usepackage{epsfig}
\usepackage{amsmath}
\usepackage{amssymb}

\newcommand{\hep}[1]{{\tt hep-ph/#1}}

\newcommand{\plb}[3]{{\it Phys.~Lett.~}{\bf B #1} #3 (#2)}

\begin{document}
\title{Pomeron - Graviton duality%
\thanks{Presented at   XLVII Cracow School of Theoretical Physics, Zakopane, Poland, June 14 - 22, 2007.}%
}
\author{Anna M. Sta\'sto
\address{Penn State University, Physics Department, 104 Davey Laboratory, University Park, PA 16802, USA\\ and \\H. Niewodnicza\'nski Institute of Nuclear Physics, Polish Academy of Science, ul.Radzikowskiego 152, 31-342 Krak\'ow, Poland}}
\maketitle
\begin{abstract}
In this lecture I give a short introduction to the high energy limit of hadronic interactions. The elements of the  Regge theory, Pomeron in  QCD and high energy scattering in  AdS/CFT correspondence 
are  presented. I discuss the resummation of the hard Pomeron which in the case of the fixed coupling leads to the value of intercept equal to two in the limit 
of the strong coupling.
\end{abstract}
  
\section{Introduction}

The high energy behavior of the  hadronic total  cross sections remains one of the biggest unsolved problems in the theory of the  strong interactions.   The problem is that, even at the  very high energies
$s\rightarrow \infty$, there is a range of scales probed in such a  process.   Many exclusive processes
with additional large scale 
can be treated using perturbative methods  thanks to  the property of  asymptotic freedom and the factorization theorems. On the other hand,  the total cross sections are notoriously difficult to evaluate from the first principles and therefore one has to rely on phenomenological models.  The high energy asymptotics of the hadronic interactions was first investigated   within the S-matrix and  Regge theory. Powerful methods based on few general principles were elaborated, despite the lack of information on the microscopic dynamics. The high energy limit in QCD was calculated within the leading logarithmic approximation in the logarithms of energy. The result was  the famous BFKL Pomeron, which indeed has the Regge behavior. More recently, in string theory the AdS/CFT conjecture opened up a new path for understanding the large coupling limit
of gauge theories. In this approach the high-energy scattering of hadrons  is dominated by the gravitational scattering with the Pomeron Regge trajectory being identified with the graviton trajectory. The picture might be complicated since  the unitarity corrections are to be taken into account, and also by the fact that the AdS/CFT conjecture is tested for UV finite and  conformal theory and so far the dual description for QCD theory is not known. In this lecture I will bring some of these ideas, namely  I will give a short and elementary introduction to the  Regge theory, high energy limit in  QCD and the strong coupling limit within the string theory.
I will also discuss the progress in resummation at high energy in QCD, which in principle allows to perform the interpolation
between small and large couplings (at least in the case of N=4 SYM theory).

\section{S-matrix and the Regge theory}
\begin{figure}[htb]
\centerline{\epsfig{file=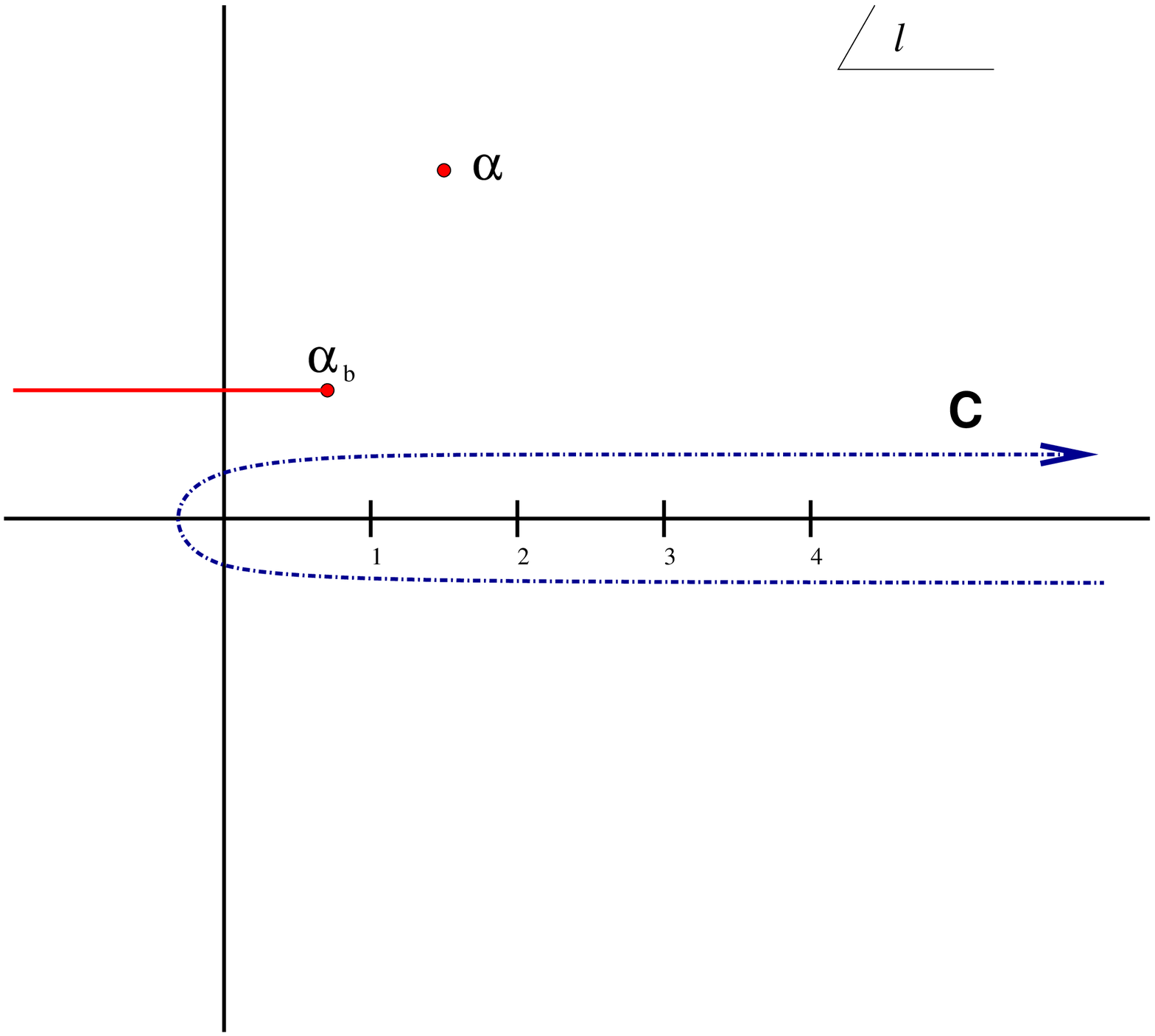,width=0.4\textwidth}\hfill\epsfig{file=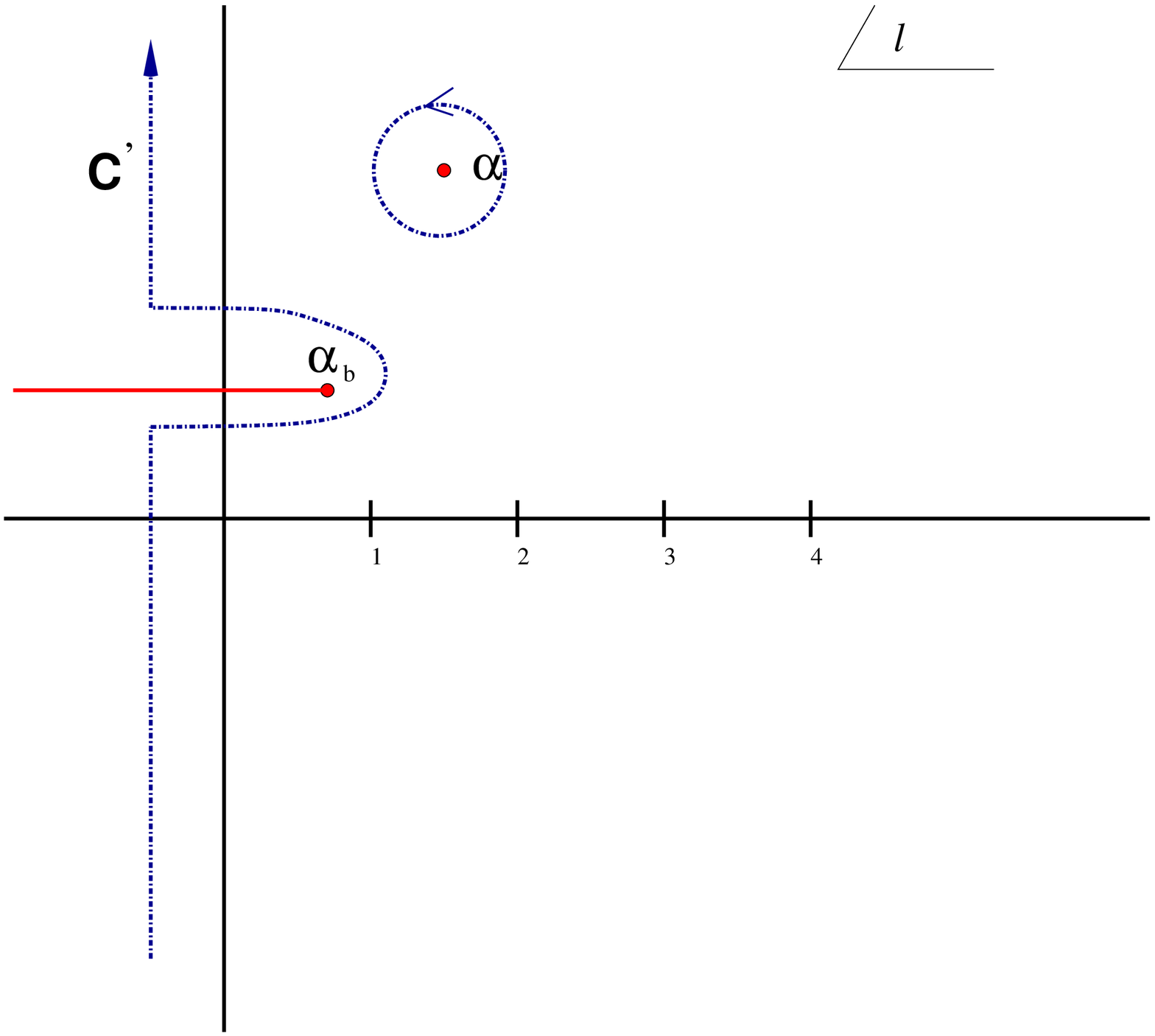,width=0.4\textwidth}}
\caption{Shape of the contour in the angular momentum plane.}
\label{fig:contourC}
\end{figure}

The S-matrix theory,  which was developed in an attempt to understand the theory of strong interactions, relied on few assumptions based on a very general and fundamental principles, see \cite{pdb_collins}. The postulates for the scattering S-matrix $\langle out| in \rangle$ were the following:
\begin{itemize}
\item Lorentz invariance. The S-matrix had to be therefore a function of the invariants $s,t,u$ and possibly masses of the incoming and outgoing particles.
\item Unitarity of $S$ matrix : $SS^{\dagger}=S^{\dagger}S=1$. The unitarity really comes from the conservation of the probability. The probability of the incoming state to scatter into a given outgoing state must be one if we sum over all possible outgoing states. 
\item Short range of the strong interactions. This allows to treat the incoming and outgoing states as free
when $t\rightarrow \infty,\; t\rightarrow-\infty$.
\item Analyticity. The $S$-matrix should be an analytic function of $s,t,u$ with only the singularities
due to stable or unstable particles and these which are required by the unitarity. This postulate is very important for the construction of the S-matrix theory but at the same time is very controversial.
\item Crossing. This is really consequence of the analyticity postulate. The physical kinematic regime for the  process
$$
a+b\rightarrow c+d \; ,
$$ is when $s>0$ and $t,u<0$. According to the analyticity postulate  the amplitude 
${\cal A}(s,t,u)_{ab\rightarrow cd}$ is an  analytic function  and therefore it can be continued to another region where $t>0$ and $s,u<0$ which gives an amplitude for a different process
$$
a+\bar{c} \rightarrow \bar{b} +d \; .
$$
Thus the same function describes both processes and one can identify 
$$
{\cal A}(s,t,u)_{a\bar{c}\rightarrow \bar{b}d}={\cal A}(t,s,u)_{ab\rightarrow cd} \; .
$$
\end{itemize}
These postulates lie at the foundations of the S-matrix approach. A particular  insight into the behaviour of the amplitude at high energy was gained by looking into its properties in the angular momentum plane. By performing  the partial wave amplitude decomposition
 for $2\rightarrow 2$ scattering in $t$-channel one can show that  the amplitude admits the representation

\begin{equation}
{\cal A}_{a\bar{c}\rightarrow \bar{b}d}(s,t) = \sum_{l=0}^{\infty} \,a_l(s) \,P_l(1+2t/s) \; ,
\label{eq:partial_wave}
\end{equation}
where $a_l(s)$ is the partial wave amplitude and the $P_l$ is  the Legendre polynomial. The continuation
to the s-channel and  the Sommerfeld-Watson transform allows to rewrite the above relation 
\begin{equation}
{\cal A}(s,t) = \frac{1}{2i} \oint_C\, dl\, (2l+1)\,\frac{a(l,t)}{\sin \pi l}\, P_l(1+2s/t) \; ,
\label{eq:SW}
\end{equation}
where now $a(l,t)$ are the functions which are analytical continuation of the amplitudes $a_l$ in (\ref{eq:partial_wave}). The contour $C$ is shown  in left plot in  Fig.~\ref{fig:contourC}, it goes around the positive real axis and encompases all the poles given by the $\sin \pi l$ in the denominator of (\ref{eq:SW}).
One can then deform the contour $C$ so that  it is parallel to the imaginary axis in $l$-plane.
There might be poles and cuts which must be encircled and so the amplitude can be rewritten as  a sum over the poles, cuts and the integral which runs along the line $(-1/2-i \infty,-1/2+i \infty)$, see right hand plot in  Fig.\ref{fig:contourC}.   We are here primarily interested in the Regge limit, i.e. in the limit when the $s \gg -t$, that is at very high energies and small angle scattering. In this limit the contribution to the amplitude is dominated by the rightmost pole in the complex angular momentum plane and the background integral over the contour $C'$, see  right hand plot in Fig.~\ref{fig:contourC}, vanishes.  In the case when the simple pole (rather than the cut) dominates, the amplitude can be approximated as
\begin{equation}
{\cal A}(s,t) \rightarrow \frac{\eta+e^{-i\pi \alpha(t)}}{2} \, \beta(t)\, s^{\alpha(t)} \; .
\label{eq:Aregge}
\end{equation}
In this equation $\alpha(t)$ is the leading Regge pole which depends on the momentum transfer $t$ and controls the high energy behavior of the amplitude; $\eta$ is the signature factor and all the normalization 
and the residue of the pole are absorbed into the  function $\beta(t)$.  
The amplitude (\ref{eq:Aregge}) can be thought of as coming from the exchange of the object-Reggeon in the $t$-channel. Its angular momentum is equal to $\alpha(t)$.  This is rather complicated object since its spin depends on $t$ and we cannot think about it as an ordinary particle since it does not have a definite representation of the Lorentz group.

So far we have considered the process with negative $t$ values but if we now look into the process with $t$-values positive then we expect the amplitude to have poles which correspond to the actual physical particles $\alpha(m_i^2)=J_i$. Here  $J_i$ is the actual   spin of physical particle with mass $m_i$.
An interesting observation made by Chew and Frautschi in the early sixties \cite{chew_frautschi} was that when  plotting
the spin of the mesons as a function of their mass the points lie on a universal straight line. This  dependence was  parametrized as 
$$
\alpha(t) = \alpha(0) + \alpha' \, t \; ,
$$
with $\alpha(0)$ being the intercept and $\alpha'$ the slope parameter.
These straight lines were called Regge trajectories.   Interestingly, the  linear behavior continues to  negative values of $t$ and it  then corresponds to the scattering process  with the exchange of the reggeon with the same quantum numbers (except spin of course, which is not defined) as the mesons lying on the trajectory. For example the process $\pi^- p\rightarrow \pi^0 n$ could be well described using the $\rho$ trajectory.  Thus the Regge trajectories turned out to be universal quantities, for positive $t$
they contain physical particles with distinct values of masses and spins, whereas for the negative $t$ values they control the energy behavior of  the scattering process.

%%%%%%%%%%%%%%%%%%%%%%%%%%%%%%%%%%%%%%%%
\subsection{The Pomeron}
The $\rho$ trajectory  had intercept $\alpha(0)<1$. From the optical theorem one obtains that the scattering cross section behaves as
$$
\sigma_{\rm TOT} \sim s^{\alpha(0)-1} \; .
$$ 
Thus the $\rho$ trajectory discussed in the previous section, which corresponds to the exchange of the object with isospin $I=1$ leads to the cross section which decreases with the energy.
Pomeranchuk showed that if there is a charge exchange in any process then the cross section would decrease at very large energies. On the other hand if the cross section increases it should be dominated by the reggeon with the quantum numbers of the vacuum. Such Reggeon is called the Pomeron. The situation could be more complicated by the Odderon state, a Reggeon which is odd under charge conjugation, whose contribution could be constant with the energy \cite{Bartels:1999yt}.

The experimental data on pp and $p\bar{p}$ scattering exhibit slow increase of the total cross section with the increasing c.m.s energy. This increase can be universally parametrized by the small power $\alpha(0)-1\simeq 0.08$ both for pp and $p\bar{p}$ collisions \cite{Donnachie:1992ny}. The  two cross sections  differ at small energies, but  they exhibit universal growth for large energies. In fact, it is very interesting that all the hadronic cross section ($pp,p\bar{p},\pi^+ p,\pi^- p, K^+ p, K^- p$)  have this universal behavior \cite{Donnachie:1992ny}. The same growth is also seen in the photoproduction cross section $\gamma p$. 
Thus we conclude that the total cross sections in strong interactions have an intriguing property of universality at high energies.
%%%%%%%%%%%%%%%%%%%%%%%%%%%%%%%%%%%%%%%%
\section{Gauge theory}
\label{sec:qcd}

\begin{figure}[htb]
\centerline{\epsfig{file=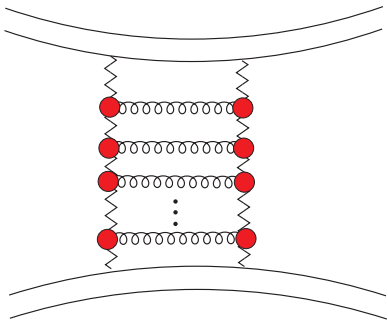,width=0.5\textwidth}}
\vspace*{-2cm}
\caption{The schematic representation of the diagram summed in the BFKL calculation. The blobs represent the effective Lipatov vertex. The gluons exchange in the $t$-channel are reggeized.
The are represented by the zigzag lines.}
\label{fig:bfkl}
\end{figure}

The $S$-matrix provided an important insight into the high energy asymptotics. It could not however
answer more detailed questions about the exact behavior since it lacked the microscopic dynamics. 
The first attempt to derive the Pomeron from QCD was done by Low and Nussinov. They considered the 2-gluon exchange process. This simple model did not have however the features expected from the Regge theory, for example it is not a Regge pole. The improved approach based on the resummation of the leading logarithms of energy was pioneered by Lipatov and collaborators \cite{Lipatov:1996ts}.  The original 2-gluon exchange model was dressed with subsequent gluon emission in the approximation $s\gg-t$. More precisely, the gluon emissions were resummed  in the limit where  each power of the strong coupling is accompanied by the logarithm of the energy.  The set of diagrams resummed in this approximation is shown in Fig.~\ref{fig:bfkl} where each gluon exchanged in  the $t$-channel acquires the 'reggeized' propagator 
$$
D^{\mu\nu}(\hat{s}_i,k_{i,T}^2)=\frac{ig_{\mu\nu}}{k_{i,T}^2} \bigg ( \frac{\hat{s}_i}{k_{i,T}^2}\bigg)^{\epsilon_G(k_{i,T}^2)} \; ,
$$
where $\hat{s}_i=(k_{i-1}-k_{i+1})^2$  with $k_i=(k_i^+,k_i^-,k_{i,T})$ being  the momenta exchanged in the ladder   and 
\begin{equation}
\epsilon_G(q_t^2)=\frac{N_c \alpha_s}{4\pi}\int d^2 k_T \frac{-q^2_T}{k_T^2 (k_T-q_T)} \; ,
\label{eq:Reggetrajectory}
\end{equation}
is the gluon Regge trajectory. The latter object was obtained by the summation of the diagrams with the virtual exchanges of gluons in the leading logarithmic approximation. As seen from (\ref{eq:Reggetrajectory}) this object is infrared divergent so formally one needs a cutoff on the small momenta to properly define it.  The vertices between the ordinary emitted gluons and the reggeized gluons are effective vertices. They   result from the summation of different tree level  single gluon emission diagrams.  The final result for the imaginary part of the amplitude with arbitrary number of the  gluon emissions is rather complicated but it turns out that it can be succinctly represented as  a solution to the  integral equation of the Bethe - Salpeter type
\begin{equation}
\omega f_{\omega}(k_{1T},k_{2T},q_T) = \delta^{(2)}(k_{1T}-k_{2T}) + \int dk'_{T}  K(k_{1T},k'_{T},q_T) \, f_{\omega}(k'_{T},k_{2T},q_T) \; .
\label{eq:bfklequation}
\end{equation}
%%%%%%%%%%%%%%%%%%%%%%%%%%%%%%%%%%%%%%%%
%%%%%%%%%%%%%%%%%%%%%%%%%%%%%%%%%%%%%%%%
Here $\omega$ is the Mellin conjugate variable to the $\ln s$ and $K$ is the (energy independent)
integral BFKL kernel which contains the real part coming  from the square of the effective vertex in Fig.~\ref{fig:bfkl} and the virtual part from the Regge trajectory. The function $f$ is called the gluon Green's function and it is dependent of the  four off-shell momenta and the rapidity (or $\omega$).
  The important property of this equation  is that it is   infrared safe, unlike the gluon Regge trajectory.

The solution to this equation was found by employing the fact that the kernel has a conformal symmetry in 2-dimensions \cite{Lipatov:1985uk}. Therefore one can diagonalize this operator with the conformal eigenfunctions. For the purposes of this lecture it is sufficient to know the solution for zero momentum transfer $t=-q_T^2=0$.
The eigenvalue equation can be written as 
$$
K \times \phi_{\nu}^n = \frac{\alpha_s N_c}{\pi} \, \chi(\nu,n)\, \phi_{\nu}^n\, ,\hspace*{1cm} \phi_{\nu}^n(k_T) = \frac{1}{\pi \sqrt{2}} (k_T^2)^{1/2+i \nu} e^{in\theta}\; ,
$$
where the eigenvalue function is 
\begin{equation}
\chi(\nu,n) = 2{\rm Re} [\psi(1)-\psi(1/2+i\nu+n/2)] \; .
\label{eq:eigenvalue}
\end{equation}
The dominant contribution is at $n=0$. The eigenvalue function  has simple poles at $$\gamma=1/2+i\nu=\dots,-2,-1,0,1,2,\dots \; ,$$ and a saddle
point at $\gamma=1/2$. 
The BFKL equation gives rise to the cut singularity which can be seen by solving 
\begin{equation}
1=\frac{\alpha_sN_c}{\pi} \frac{1}{\omega} \chi(n=0,\gamma) \; ,
\label{eq:polebfkl}
\end{equation}
for $\omega$. The cut structure is clear since as $\gamma$ varies along the imaginary axis $(1/2-i\infty,1/2+\infty)$ the value of $\omega$ from this equation varies from $-\infty$ to $4\ln2 \alpha_s N_c/\pi$.

One can also find the solution by the saddle point method. To this aim one can expand the kernel around the saddle point $\nu=0$ to get  $$\chi(\nu)\simeq 4\ln2 -14 \zeta(3) \nu^2 \; .$$ This leads to the following solution in the diffusion approximation
\begin{multline}
f(\ln s/s_0,k_{1T},k_{2T}) \simeq \\ \simeq {\cal N}(\alpha_s,s,k_{1T},k_{2T}) \bigg( \frac{s}{s_0}\bigg)^{\omega_0} \exp\bigg(-\frac{\pi \ln ^2 \frac{k_{1T}^2}{k_{2T}^2}}{28 \zeta(3) \alpha_s N_c \ln s/s_0}\bigg) \; ,
\label{eq:bfklsol_diff}
\end{multline}
where normalization function ${\cal N}$ depends on the energy and momenta but the leading behavior has been factored out.
We see that the energy dependence of the solution is governed by the power behavior with the power equal to the value at the minimum of the kernel $\omega_0=4\ln 2 \frac{\alpha_s N_c }{\pi}$. The last term on the right hand side of Eq.~(\ref{eq:bfklsol_diff})
is the diffusion term. The transverse momenta play the role of the coordinates and the logarithm of the energy is like the imaginary time. The diffusion in the transverse momenta is then controlled by the second derivative of the kernel around its minimum.  Therefore the BFKL resummation of the leading logarithms in the energy showed that the gluons 'reggeize' i.e. they form composite objects at high energy and that the amplitude is dominated by the Regge cut.
Unfortunately the BFKL leading logarithmic resummation turned out to be incompatible with the experimental data. The power behavior $s^{\omega_0}$ with $\omega_0=4 \ln2 \frac{\alpha_s N_c}{\pi}\simeq 0.5$ (say for typical values of  $\alpha_s\simeq 0.2$) is much too strong not only for the total proton-proton cross sections but also for the growth of the structure function $F_2$ in Deep Inelastic Scattering  of electron on a proton target where the behavior is roughly  $F_2(x) \sim x^{-\lambda_{\rm eff}}\, , \;  \lambda_{\rm eff}=0.2-0.3$\footnote{We mean here that the effective behavior can be parametrized by the power of this value. The data  are very well described by the conventional renormalization group equations which do not posses this type of singularity.}.  Therefore it became clear that there is a need for higher order terms. We will come back to this problem in Sec.~\ref{sec:resum}.

%%%%%%%%%%%%%%%%%%%%%%%%%%%%%%%%%%%%%%%%
\section{Graviton and string theory in  $AdS_5$ background}
\label{sec:string}

The graviton is thought to be a quantum of the gravitational field and, if it exists, it must be a particle of spin two, see for example \cite{Veltman:1975vx}.
Since it couples to energy-momentum tensor it cannot be a scalar. It cannot be a vector particle also,
since it would lead to difference between particles and antiparticles which contradicts the
experiments. It has to be massless object since the gravity is a long range force. The universality of its couplings to particles can be shown by analyzing the amplitudes for the emissions of soft gravitons and employing Ward identities \cite{Weinberg:1964ew}. Then directly
from the condition of energy-momentum conservation it follows that all the couplings of gravitons to particles are equal. Therefore the principle of equivalence is a natural consequence of the Lorentz invariance for the massles spin 2 particles.   In string theory the graviton emerges as a particular
closed string state.   
The AdS/CFT conjecture gives a tool for analyzing the gauge theory in a regime where the standard perturbative methods are insufficient. It states that the two theories: conformal field theory in $d=D-1$ dimensions
and the string theory in an anti de Sitter space-time in $D$ dimensions are related to each other, \cite{Maldacena:2003nj}. More precisely it states that, the limits of these two different theories which contain different degrees of freedom
are interchanged when the coupling $g^2 N$ is varied. When the coupling $g^2 N_c \gg 1$ then the gauge theory is strongly coupled but the string theory is weakly coupled. On the other hand when $g^2 N \ll 1$ the gauge theory 
is weakly coupled, but the gravity is strongly coupled. The conjecture relates the boundary values of the  fields on the gravity side to the local operators on the gauge theory side.
The correspondence was checked  in a particular case of the conformal field theory $N=4$ super Yang-Mills. This theory apart from the gauge field $A_{\mu}$ contains also six scalar fields $\phi_i$
and four fermions $\chi_j$. All the fields transform in the adjoint representation. The theory is UV finite and the coupling does not run, this fact makes this theory quite different from the  QCD.
Nevertheless, the infrared regime is similar to QCD and since the computations are easier in this theory it can be thought of as a useful laboratory for QCD.

The high energy limit of the scattering amplitudes was investigated in the gravity dual, and it turned out that the exchange would be dominated by the  graviton state with $j_0=2$ \cite{Janik:1999zk,Brower:2006ea}.
What is also interesting that the same diffusion pattern was found for the amplitude as in the weak coupling limit. This was interpreted as a diffusion in the fifth(radial) coordinate of AdS space
and on the gauge theory side this corresponds to the diffusion in the transverse momenta along the ladder.
The only difference is in the value of the power and the diffusion coefficient
\begin{eqnarray}
j_0= \omega_0+1& = & 2-\frac{2}{\sqrt{g^2 N}}, \;  \;{\cal D}=\frac{1}{2\sqrt{g^2 N}},  \; \; g^2 N \gg 1 \label{weak} \\
j_0=\omega_0+1& = & 1+4 \ln2 \frac{\alpha_s N}{\pi},  \; \; {\cal D}=7 \zeta(3) \frac{\alpha_s N_c}{\pi}, \; \;   g^2 N \ll 1
\label{strong}
\end{eqnarray}
where $j_0=\omega_0+1$ with $\omega_0$ from the previous notation.
 At small values of the coupling we have a linear increase with the coupling according to the leading logarithmic
approximation (\ref{weak}). At large values of the coupling the intercept becomes exactly $2$ with the correction that vanishes as $1/\sqrt{g^2 N}\;\;$ (\ref{strong}).
We see that we have two results which should be good approximations to two different regions of the coupling.
The problem is that they are totally disconnected from each other and it is hard to see that they actually describe the same object.

%%%%%%%%%%%%%%%%%%%%%%%%%%%%%%%%%%%%%%%%
\section{Resummation at small $x$}
\label{sec:resum}

The leading logarithmic approximation gave a very large value for the intercept of the Pomeron (\ref{weak}). Assuming the typical value of the coupling of about $0.2$, the Pomeron
intercept value from this calculation is about $0.5$. This is in a blatant disagreement with the experimental data, especially the structure function data   in deep inelastic scattering. 
The next-to-leading correction \cite{NLLx} turned out to be very large, 
$$
j_0=1+4 \ln 2 \frac{\alpha_s N_c}{\pi} (1-6.45 \frac{\alpha_s N_c}{\pi})  \;. 
$$
Therefore it became immediately clear that there is a need for the resummation of this series. There are several sources of very large corrections. The first of them is the running coupling. 
It is  fixed in the leading logarithmic calculation due to the subleading contribution (from the point of view of the leading logarithms in energy) from the gluon loops. 
It starts to run only at the next to leading level (NLLx).  The other important  corrections include the kinematical constraint and the requirement of the energy momentum-conservation.
 This was shown \cite{Kwiecinski:1996td} to give important contribution
even before the explicit NLLx calculation.  Finally, there are also  corrections coming from   the quarks in the evolution. 
Here we will only consider the corrections which come from the kinematics since they are common to both QCD and $N=4$ SYM theory. The kinematical constraint
comes from a more careful treatment of the final state phase space. The leading logarithmic approximation is done both on the level of the amplitude and on the phase space
of emitted gluons. A careful analysis shows that, there is a region of momenta for which the emitted gluons are off-shell. The kinematic constraint imposed
onto the real emission part of the kernel corrects this problem. The result is an all-order resummation of the subleading terms. In particular it was shown that this 
constraint is responsible for the triple collinear poles which appear in the next-to-leading calculation and which constitute numerically a large part of the corrections \cite{Salam:1998tj}.
It turns out that it is still insuficient, since the energy momentum is not conserved exactly. Various schemes were proposed, \cite{resum}  here we will consider a very simple
model which has energy-momentum conservation imposed on the level of the eigenvalue \cite{ams}. It is rather brute-force method but it does give qualitatively results
which are expected from the gravity calculation at strong coupling.  
The anomalous dimensions in the usual renormalization group approach have a constraint that 
\begin{eqnarray}
\gamma_{gg}(j=2)+2 N_f \gamma_{qg}(j=2) & = & 0 \; , \nonumber \\
\gamma_{gq}(j=2)+ \gamma_{qq}(j=2) & = & 0 \; ,
\label{eq:emco}
\end{eqnarray}
which is independent of the order of perturbation theory.  In $N=4$ SYM the condition is much simpler 
$$\gamma_{\rm uni}(j=2)=0 \; ,$$
where $\gamma_{\rm uni}$ is defined for example in \cite{Kotikov:2003fb,Kotikov:2004er}.
One can evaluate the anomalous dimension from the BFKL calculation by solving the equation (\ref{eq:polebfkl}) for $\gamma$. This anomalous dimension does not satisfy the energy momentum constraint.
This is due to the fact, that as mentioned above, the approximations are made on the level of the amplitude and on the level of the phase space integral.
The simple model that satisfies the energy momentum conservation was taken in \cite{ams} simply as
\begin{eqnarray}
1& =& \bar{\alpha}_s \, \gamma_{gg}(\omega) \, \chi(\omega,\gamma)  \; ,\nonumber \\
\chi({\omega,\gamma}) & =&  -2 \gamma_E - \psi(\gamma+\omega/2)-\psi(1-\gamma+\omega/2)  \; .
\label{eq:simple_model}
\end{eqnarray}
The shifts in the arguments of the kernel eigenvalue come from the kinematical constraint. The anomalous dimension in front of the eigenvalue guarantees the energy momentum conservation
when $j=\omega+1=2$. The multiplicative model above is probably too naiive. Nevertheless it gives the result that the intercept becomes $2$ for large values of the coupling $\alpha_s$, see also \cite{Kotikov:2003fb,Kotikov:2004er}. 

\begin{figure}[htb]
\centerline{\epsfig{file=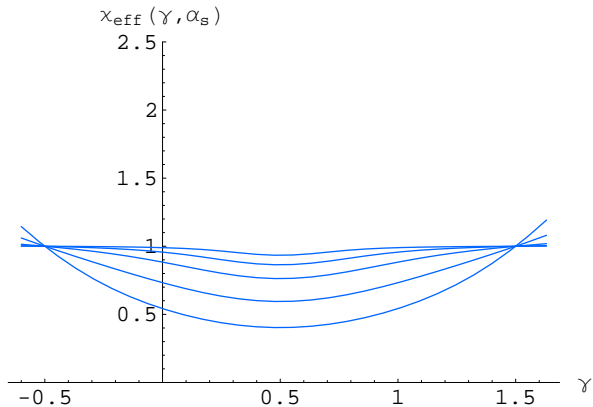,width=0.7\textwidth}}
\caption{The solution for $\omega$ from Eq.~(\ref{eq:simple_model}). Fixed points result from the energy momentum constraint.}
\label{fig:eigenvalue}
\end{figure}

\begin{figure}[htb]
\centerline{\epsfig{file=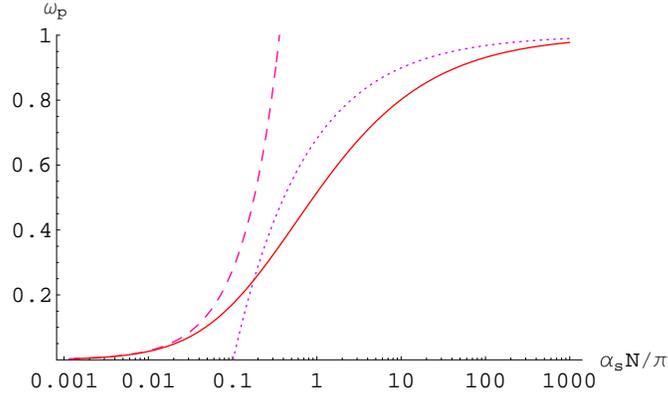,width=0.7\textwidth}}
\caption{The value of the intercept, calculated from the minimum of the resummed eigenvalue as a function of the coupling constant $\alpha_s N_c/\pi$.}
\label{fig:intercept}
\end{figure}

Solving this equation for $\omega$ gives the result shown in Fig.~\ref{fig:eigenvalue}. We see that the constraint forces the curve to have to fixed points. Unlike the leading logarithmic case, where the kernel eigenvalue can take arbitrary values for the 
large values of the coupling constant the minimum of this kernel $j_0=1+\omega_0$ is constrained to the interval $[0,2]$. The first correction goes as $1/\sqrt{\alpha_s}$ at large values of the coupling, compare (\ref{strong}).
The second derivative goes as $1/\alpha_s$ which is probably an artefact of the particular multiplicative simple model. One can evaluate the minimum of this eigenvalue as a function of the coupling constant,
which is shown in Fig.~\ref{fig:intercept}. We see that the model provides a very nice interpolation between the small and large values of the coupling.

\begin{figure}[htb]
\centerline{\epsfig{file=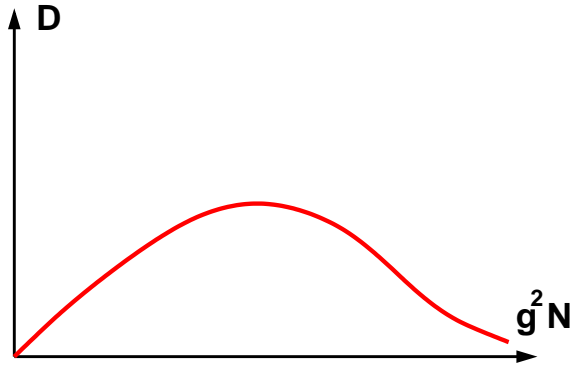,width=0.6\textwidth}}
\caption{The value of the diffusion coefficient as a function of the coupling constant.}
\label{fig:diffusion}
\end{figure}

The other interesting feature is the behavior
of the diffusion pattern in weak and strong coupling limits. From (\ref{weak},\ref{strong}) we see that the diffusion vanishes both at weak and at strong values of the coupling. It is also clear from Fig.~\ref{fig:eigenvalue}. The vanishing
at small coupling is clear, since it is proportional to the coupling. At strong coupling the eigenvalue becomes very flat as it tends ot a constant. The second derivative then vanishes in this limit.
The physical interpretation is that this region is dominated by the soft gluons with vanishing energy. The qualitative behavior of the diffusion parameter as a function of the coupling is shown in Fig.~\ref{fig:diffusion}.
It is zero at $\alpha_s N_c=0,\alpha_s N_c=\infty$ and it has to have a maximum at some intermediate values of $\alpha_s N_c$.
%%%%%%%%%%%%%%%%%%%%%%%%%%%%%%%%%%%%%%%%
%%%%%%%%%%%%%%%%%%%%%%%%%%%%%%%%%%%%%%%%

%%%%%%%%%%%%%%%%%%%%%%%%%%%%%%%%%%%%%%%%
\section{Conclusions}
In these lectures I  gave a brief overview of the high energy limit in hadronic collisions.  It is expected that the high energy limit is governed by the exchange of the
object with the quantum numbers of the vacuum, called the Pomeron.  In QCD it can be calculated by the summation of the Feynman diagrams in the leading
logarithmic approximation in the logarithms of the energy. The result leads to the very strong increase of the amplitude with the energy and  it is not
compatible with the experimental data. The resummation of the subleading corrections  was shown to tame this rapid growth and reduce the value of the intercept of the Pomeron.
The large amount of the corrections comes from the exact treatment of the kinematics: energy-momentum conservation constraint and the kinematical constraint.
By putting these two constraints onto the kernel, one can show that there is  a limit on a value of the Pomeron intercept when the coupling constant becomes infinite.
It corresponds to $\omega_0=1$ which is the value  if there was an exchange of an object with spin two. Several important questions remain.
The unitarity corrections should become equally important in addition to the single Pomeron exchange, \cite{Hatta:2007he}. The graviton itself emerges here as an object
which consists of very soft gluons , in the limit where the infrared divergences of the gauge theory cancel. The considerations so far were only done at the level
of the fixed coupling in a model which is close to N=4 SYM theory rather than QCD. Running coupling effects and mixing with quarks must be taken into 
account when considering real QCD.

%%%%%%%%%%%%%%%%%%%%%%%%%%%%%%%%%%%%%%%%
\section*{Acknowledgments}
I would like to thank the organizers of the Cracow School of Theoretical Physics for a possibility to give this presentation and for the
very interesting school.
This research is supported  by the U.S. D.O.E. under grant
number DE-FG02-90ER-40577 and by the 
 Polish Committee for Scientific
Research grant No. KBN 1 P03B 028 28. 
%%%%%%%%%%%%%%%%%%%%%%%%%%%%%%%%%%%%%%%%

\end{document}